\documentclass[11pt]{article}

\usepackage{iftex}
\ifPDFTeX\usepackage[utf8]{inputenc}\fi
\usepackage[T1]{fontenc}
\usepackage{lmodern}
\usepackage[a4paper,margin=1in]{geometry}
\usepackage{amsmath,amssymb,amsthm,mathtools,bm}
\usepackage{graphicx}
\usepackage{microtype}
\usepackage{xcolor}
\usepackage[font=small,labelfont=bf]{caption}
\usepackage[section]{placeins}
\usepackage{xurl}
\usepackage{hyperref}

\definecolor{linkblue}{HTML}{326B83}

\hypersetup{
  colorlinks=true,
  linkcolor=linkblue,
  citecolor=linkblue,
  urlcolor=linkblue
}

\graphicspath{{./}}

\newcommand{\R}{\mathbb{R}}
\newcommand{\1}{\mathbf{1}}
\newcommand{\Weff}{W_{\mathrm{eff}}}
\newcommand{\KE}{K_{E}}
\newcommand{\Aop}{\mathcal{A}}
\newcommand{\dd}{\,\mathrm{d}}

\newcommand{\figimg}[2][0.97\linewidth]{\includegraphics[width=#1]{#2}}

\title{\textbf{Localization Landscapes as Operator-Induced Geometry for Heterogeneous Systems}}

\author{
Josué García-Ávila\\
Department of Mechanical Engineering\\
Columbia University, New York City, NY 10027, USA\\
\href{mailto:jg5020@columbia.edu}{\texttt{jg5020@columbia.edu}}
}
\date{}

\begin{document}
\maketitle

\begin{abstract}
Heterogeneous operators often organize response through hidden barriers, wells, and weakly communicating compartments that are not faithfully described by Euclidean distance or raw graph connectivity alone. We study this structure through the localization landscape, obtained from a single source solve, and the associated effective potential \(\Weff = 1/u\). Starting from both discrete Schr\"odinger-type operators and heterogeneous reaction--diffusion operators, we show how \(\Weff\) induces effective wells, basin partitions, and Agmon-type neighborhoods that serve as an operator-aware geometry for localization and transport. This geometry explains the spatial support of low-energy eigenmodes, produces adaptive receptive fields on a fixed graph, sharpens locality in attention-like interactions, and yields bottleneck-sensitive modulation of diffusion. We also include a finite-element reaction--diffusion example with strongly compartmentalized coefficients, demonstrating that the same framework extends beyond simple grid operators to heterogeneous continuum systems. The results support a precise conclusion: localization-landscape geometry is a strong inductive bias when relevance is controlled by confinement, barriers, and low-energy accessibility, but its benefits are task-dependent rather than universal.
\end{abstract}


\section{Introduction}

Many heterogeneous differential and graph-based operators exhibit behavior that is organized not by physical proximity alone, but by hidden energetic barriers and preferred subregions. In such settings, low modes localize inside effective wells, propagation is redirected through narrow passages, and interactions across barriers are strongly attenuated. Classical geometric notions based solely on Euclidean distance, mesh adjacency, or hop count can therefore be poorly aligned with the actual low-energy structure of the operator. A more faithful description is obtained by endowing the domain with a geometry induced by the operator itself rather than by the ambient discretization alone \cite{Filoche2012,Arnold2016,Arnold2019Effective,David2021}.

The central object in localization-landscape theory is the positive solution \(u\) of a source problem of the form
\begin{equation}
Lu = 1,
\label{eq:landscape_generic}
\end{equation}
for a suitable positive operator \(L\). Since the original work of Filoche and Mayboroda, the landscape has been shown to predict the support of localized low-energy states and to reveal the hidden partition of disordered media into weakly communicating subregions \cite{Filoche2012}. A major refinement is that the reciprocal field \(1/u\) acts as an effective confining potential, sharply encoding wells, barriers, and tunneling bottlenecks \cite{Arnold2016,Arnold2019Effective}. This view yields decay estimates, spectral surrogates, and computationally efficient approximations of localization structure without repeated eigenvalue solves \cite{Arnold2019Computing,David2021}.

The original landscape framework was introduced for disordered continuous operators, while later work clarified the role of \(1/u\) as an effective confining potential and extended the same perspective to discrete symmetric \(M\)-matrices \cite{Filoche2012,Arnold2016,Arnold2019Effective,Filoche2021Mmatrix}. The same viewpoint has also proved useful in disorder-aware transport models for semiconductors \cite{Filoche2017I,Piccardo2017II,Li2017III}, in graph and quantum-graph settings related to Agmon-type decay \cite{Harrell2018QuantumGraphs,Steinerberger2023Graphs}, and in recent extensions to highly excited and interacting systems \cite{Herviou2020,Stellin2023}. These developments suggest that localization-landscape geometry is not merely an interpretive tool, but a practical operator-aware surrogate for the structures that govern response, transport, and spectral organization in heterogeneous systems.

From the computational side, graph and manifold methods have long emphasized that Laplacians, diffusion, and spectral embeddings can reveal nontrivial organization beyond raw coordinates \cite{Fiedler1973,Belkin2003,Coifman2006,ShiMalik2000,VonLuxburg2007,Hammond2011,Shuman2013,Bronstein2017}. In parallel, modern learning architectures rely heavily on message passing, diffusion, and attention-based interactions \cite{Gilmer2017,Gasteiger2019,Shaw2018}. This motivates a natural question: can localization-landscape geometry provide an interpretable, operator-aware prior for downstream graph and PDE computations, especially in settings where barrier structure matters more than bare adjacency?

In this work, we develop that viewpoint in two settings: a discrete two-well operator on a regular grid and a finite-element reaction--diffusion operator with highly heterogeneous coefficients. For each case, we compute the landscape \(u\), define the effective potential \(\Weff = 1/u\), extract wells and basin partitions, and build Agmon-type neighborhoods. We then test this geometry in a sequence of increasingly downstream tasks: explaining low-mode localization, defining adaptive receptive fields, modulating diffusion across a bottleneck, biasing attention toward block-local interactions, and examining geometry-guided sparsification. The contribution is intentionally nuanced. Our aim is not to claim that landscape geometry improves every task uniformly, but rather to identify where the geometry is genuinely informative and where its transfer is mixed.

\section{Operator-aware landscape geometry}

\subsection{Discrete and continuum operators}

For the discrete example, we consider a symmetric positive operator
\begin{equation}
A = L + \operatorname{diag}(V),
\label{eq:discrete_operator}
\end{equation}
where \(L\) is the five-point grid Laplacian and \(V\) is a heterogeneous potential defining two favorable regions separated by a high vertical barrier. The localization landscape is the vector \(u \in \R^{n}\) solving
\begin{equation}
A u = \1,
\label{eq:discrete_landscape}
\end{equation}
with \(\1\) the all-ones vector. When \(A\) is a positive symmetric \(M\)-matrix, one has \(u_i > 0\), so the entrywise reciprocal
\begin{equation}
\Weff(i) = \frac{1}{u_i}
\label{eq:discrete_weff}
\end{equation}
is well defined and acts as an effective potential \cite{Arnold2016,Arnold2019Effective,Filoche2021Mmatrix}.

For the continuum example, we consider a heterogeneous reaction--diffusion operator
\begin{equation}
\Aop := -\nabla \cdot \bigl(k(x)\nabla (\cdot)\bigr) + \sigma(x)\,(\cdot),
\label{eq:rd_operator}
\end{equation}
with homogeneous Dirichlet conditions on the three interior cooling holes and homogeneous natural Neumann conditions on the outer boundary, as in the finite-element implementation. Here \(k(x)\) is a spatially varying conductivity, \(\sigma(x)\) is a reaction or uptake field, and \(q(x)\) is a localized source. The physical field \(T\) solves
\begin{equation}
\Aop T = q(x)
\quad \text{in } \Omega.
\label{eq:rd_pde}
\end{equation}
From the same operator we compute the landscape \(u\) by solving
\begin{equation}
-\nabla \cdot \bigl(k(x)\nabla u(x)\bigr) + \sigma(x)\,u(x) = 1
\quad \text{in } \Omega,
\label{eq:rd_landscape}
\end{equation}
with the same boundary conditions. Throughout, we assume \(k(x)\ge k_0>0\) and \(\sigma(x)\ge 0\), with a nonempty Dirichlet portion supplied by the cooling holes, giving a coercive mixed-boundary problem. Under these boundary conditions, the landscape \(u\) satisfies \(u>0\) in the interior of \(\Omega\) and vanishes on the Dirichlet boundary. Accordingly, \(\Weff=1/u\) is well defined in the interior and becomes large near Dirichlet boundaries. The effective potential is again defined as
\begin{equation}
\Weff(x) = \frac{1}{u(x)}.
\label{eq:continuous_weff}
\end{equation}

\subsection{Effective wells, basins, and Agmon geometry}

Given an energy threshold \(E\), we define the effective well set in the discrete case by
\begin{equation}
\KE := \{\, i : \Weff(i) \le E \,\},
\label{eq:well_set_discrete}
\end{equation}
and analogously in the continuum case by the sublevel set
\begin{equation}
\KE := \{\, x \in \Omega : \Weff(x) \le E \,\}.
\label{eq:well_set_continuum}
\end{equation}
Low values of \(\Weff\) identify favorable localization regions, while high values define energetic barriers.

To quantify accessibility, we use an Agmon-type distance built from the positive part of \(\Weff-E\). For scalar conductivity \(k(x)\), the continuum reference geometry is
\begin{equation}
d_E(x,y)=\inf_{\gamma:x\rightsquigarrow y}
\int_0^1 \sqrt{\frac{(\Weff(\gamma(s))-E)_+}{k(\gamma(s))}}
\,\|\gamma'(s)\|\,\dd s,
\label{eq:agmon_continuum}
\end{equation}
where the infimum is over piecewise \(C^1\) paths. The unit-conductivity case reduces to the usual effective-potential integrand. In both numerical implementations we use the same discrete surrogate. For every nonzero off-diagonal coupling \(a_{ij}\), define
\begin{equation}
v_i=(\Weff(i)-E)_+,\qquad
\bar v_{ij}=\tfrac12(v_i+v_j),\qquad
c_{ij}=\max\!\left\{10^{-9},\,
\log\!\left(1+\sqrt{\frac{\bar v_{ij}}{\max(|a_{ij}|,10^{-12})}}\right)\right\}.
\label{eq:agmon_edge_cost}
\end{equation}
The graph distance is the shortest accumulated edge cost,
\begin{equation}
\rho_E(i,j)=\inf_{\gamma:i\rightsquigarrow j}\sum_{\ell=0}^{m-1}
c_{\gamma_\ell\gamma_{\ell+1}}.
\label{eq:agmon_discrete}
\end{equation}
The positive floor preserves intra-well edges during sparse symmetrization. Basin assignment uses the nearest well in this graph distance. This log-cost is a discrete accessibility surrogate; we do not assert mesh convergence to \eqref{eq:agmon_continuum} or identify its radius with a continuum length. The grid threshold \(E\) is selected from low spectral values and low quantiles of \(\Weff\); the finite-element threshold uses the prescribed low-percentile well-selection rule. Both prefer a small number of sufficiently large connected well components.

\subsection{\texorpdfstring{Why \(1/u\) acts as an effective potential}{Why 1/u acts as an effective potential}}

The interpretation of \(1/u\) is not merely heuristic. For a Schr\"odinger-type eigenproblem
\begin{equation}
(-\Delta + V)\phi = \lambda \phi,
\label{eq:schrodinger_eig}
\end{equation}
together with the landscape equation
\begin{equation}
(-\Delta + V)u = 1,
\label{eq:schrodinger_landscape}
\end{equation}
one may write \(\phi = u\psi\). A direct calculation yields
\begin{equation}
-\frac{1}{u^{2}} \nabla\cdot\bigl(u^{2}\nabla \psi\bigr) + \frac{1}{u}\psi = \lambda \psi.
\label{eq:effective_potential_transform}
\end{equation}
Hence \(1/u\) appears explicitly as the confining term in the transformed equation, which explains why it captures localization geometry more faithfully than the raw potential in many disordered settings \cite{Arnold2016,Arnold2019Effective}. In the discrete case, analogous estimates remain valid for general symmetric \(M\)-matrices, yielding exponential localization bounds in terms of the corresponding Agmon distance \cite{Filoche2021Mmatrix,Steinerberger2023Graphs}.

\section{Discrete two-well example}

\subsection{Landscape recovery of wells and barriers}

\begin{figure}[htbp]
\centering
\figimg{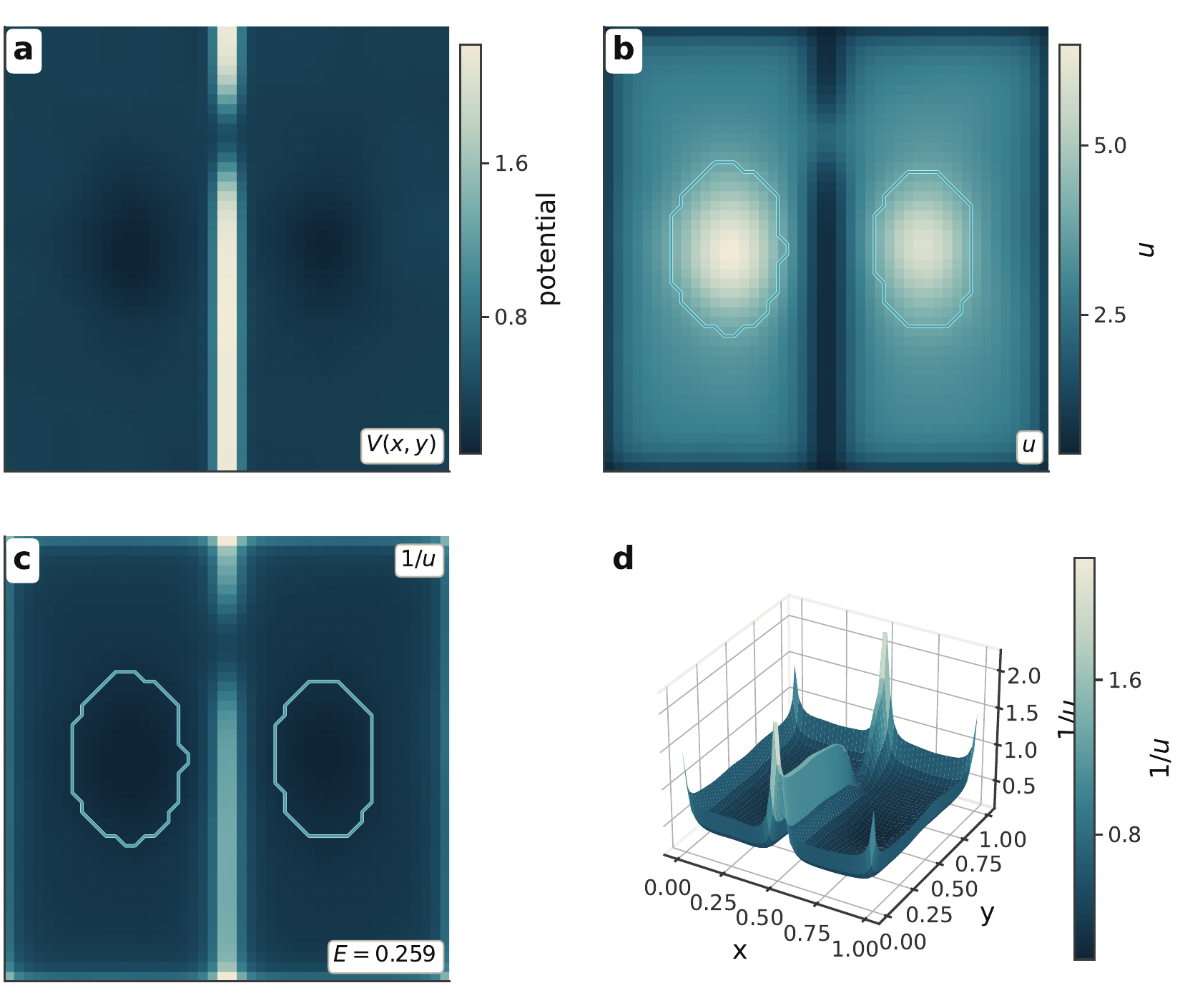}
\caption{Landscape geometry in the discrete two-well operator. (a) Input potential \(V\), including the central barrier and its partial opening. (b) Landscape \(u=A^{-1}\1\). (c) Effective potential \(1/u\), with \(E=0.259\) selecting two connected wells. Blue-turquoise contours in (b,c) mark their boundaries. (d) Surface representation of the same \(1/u\) field. Panels (c,d) share a single color scale. Low effective potential identifies accessible wells, while the ridge and domain boundary identify barriers.}
\label{fig:twowell}
\end{figure}

We begin with the two-well discrete operator \eqref{eq:discrete_operator}--\eqref{eq:discrete_landscape}. The imposed potential consists of two low-potential regions separated by a strong vertical barrier with a partial opening in the upper portion of the domain. The resulting landscape geometry is summarized in Fig.~\ref{fig:twowell}. The input potential in Fig.~\ref{fig:twowell}(a) defines the coarse arrangement of favorable and unfavorable regions, but the landscape in Fig.~\ref{fig:twowell}(b) already reveals operator-induced compartments through the elevated response inside the wells and suppression across the central barrier. The reciprocal landscape in Fig.~\ref{fig:twowell}(c) sharpens this organization into two connected low-energy basins; in the present example, the threshold \(E=0.259\) isolates two components of \(\KE\). The surface view in Fig.~\ref{fig:twowell}(d) makes the induced geometry particularly transparent: two broad basins are separated by a high ridge, while the domain boundary also acts as an elevated region.

These observations matter because they show that the relevant geometry is neither a crude thresholded partition nor a restatement of the raw potential. Instead, \(\Weff\) encodes a full scalar geometry whose wells and ridges organize localization, transport, and communication across the domain. This is precisely the type of operator-aware structure that classical hop distance or Euclidean proximity fail to capture.

\subsection{Low-energy eigenmodes follow the effective wells}

\begin{figure}[htbp]
\centering
\figimg{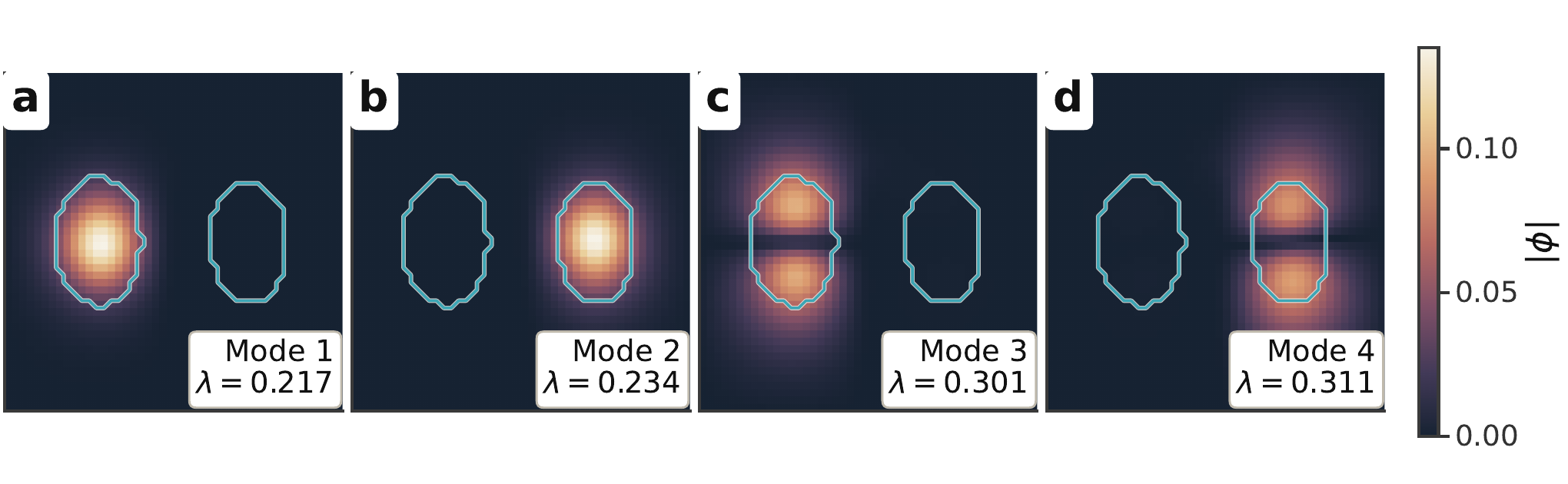}
\caption{Magnitudes of the first four low-energy eigenmodes of \(A=L+\operatorname{diag}(V)\). The panels share one amplitude scale; blue-turquoise contours show the landscape wells. The first two modes concentrate in opposite wells, while the next two retain well-scale confinement and develop internal nodal structure.}
\label{fig:eigenmodes}
\end{figure}

The spectral evidence is shown in Fig.~\ref{fig:eigenmodes}. The first four eigenmodes of \(A=L+\operatorname{diag}(V)\) are displayed on top of the well structure extracted from \(1/u\). The first two modes, shown in Fig.~\ref{fig:eigenmodes}(a,b), are strongly localized in the left and right wells, respectively, consistent with the two disconnected low-energy regions identified in Fig.~\ref{fig:twowell}(c). The higher low-energy modes in Fig.~\ref{fig:eigenmodes}(c,d) remain confined to the wells and develop internal nodal structure rather than delocalizing across the barrier.

This is precisely the behavior expected from a localization-landscape interpretation: the wells extracted from \(\Weff\) act as the dominant compartments for low-energy eigenstates. In this sense, the landscape does not merely correlate with the spectrum qualitatively; it recovers the correct large-scale partition of the low-energy eigenspace. That computational asymmetry is one of the strongest practical advantages of the framework \cite{Arnold2019Computing}.

\subsection{Adaptive receptive fields on the same graph}

\begin{figure}[htbp]
\centering
\figimg{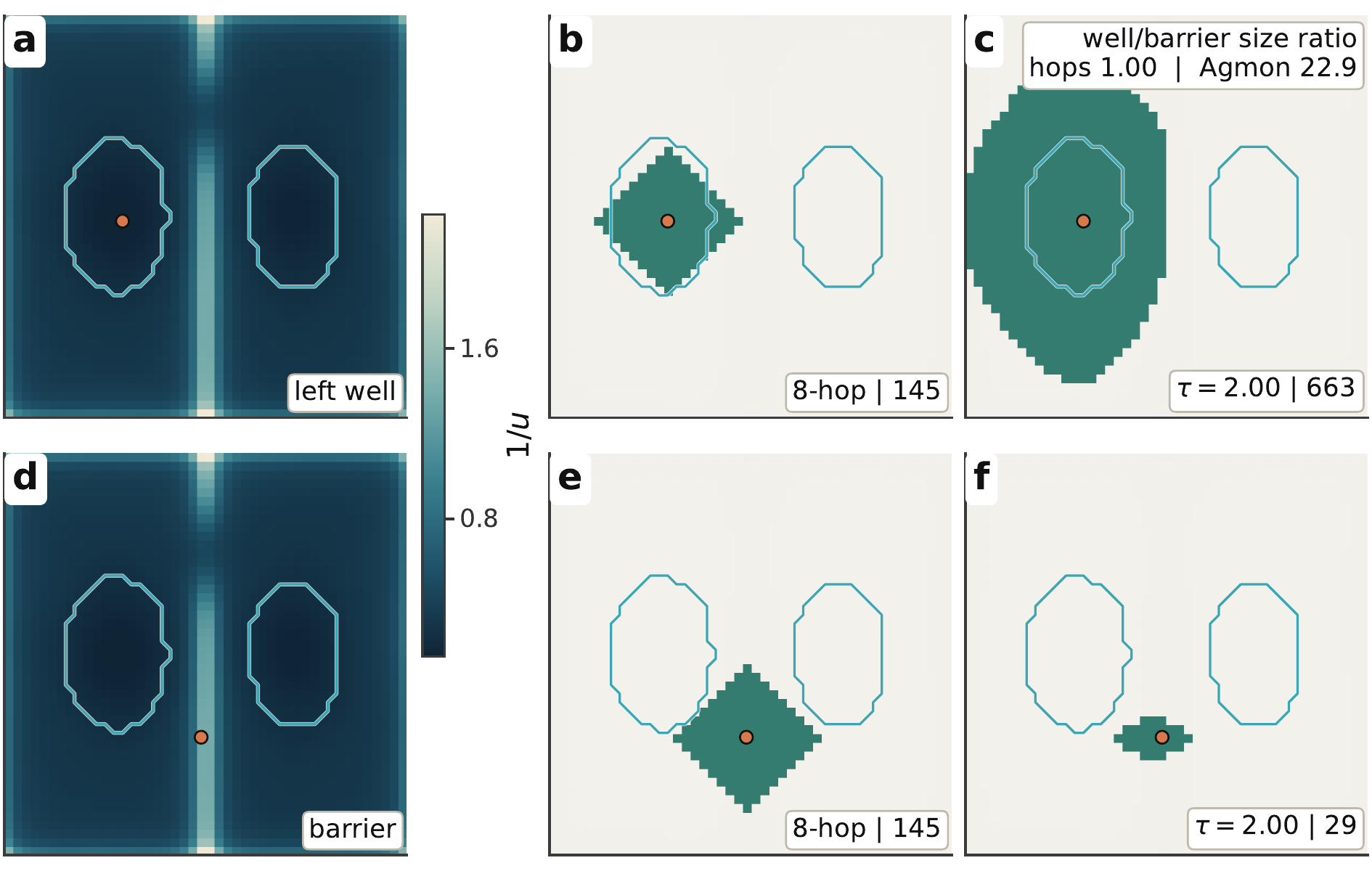}
\caption{Adaptive receptive fields on the same graph. (a,d) Interior seeds in the left well and on the central barrier, respectively. (b,e) Both 8-hop balls contain \(145\) nodes because neither is clipped by the domain boundary. (c,f) At \(\tau=2.0\), the Agmon balls contain \(663\) and \(29\) nodes, respectively. The corresponding size ratio is \(22.86\). Blue-turquoise contours identify the wells and warm markers identify the seeds.}
\label{fig:receptive}
\end{figure}

A first downstream test is receptive-field construction on a fixed graph. Both seeds in Fig.~\ref{fig:receptive} are chosen at least nine grid points from the domain boundary. Their 8-hop neighborhoods therefore contain exactly \(145\) nodes each; boundary clipping does not explain the contrast. With the same Agmon radius \(\tau=2.0\), the left-well seed reaches \(663\) nodes and the barrier seed reaches \(29\), a well-to-barrier size ratio of \(22.86\). Thus the adaptive difference arises from the operator geometry on an unchanged graph.

This result is conceptually important. It shows that the same underlying graph can induce radically different neighborhoods once accessibility is measured through operator geometry rather than through raw topology. In problems where relevance is controlled by barriers and confinement, operator-aware receptive fields are therefore much more informative than a fixed hop-radius construction.

\subsection{Landscape-gated diffusion is bottleneck-sensitive}

\begin{figure}[htbp]
\centering
\figimg{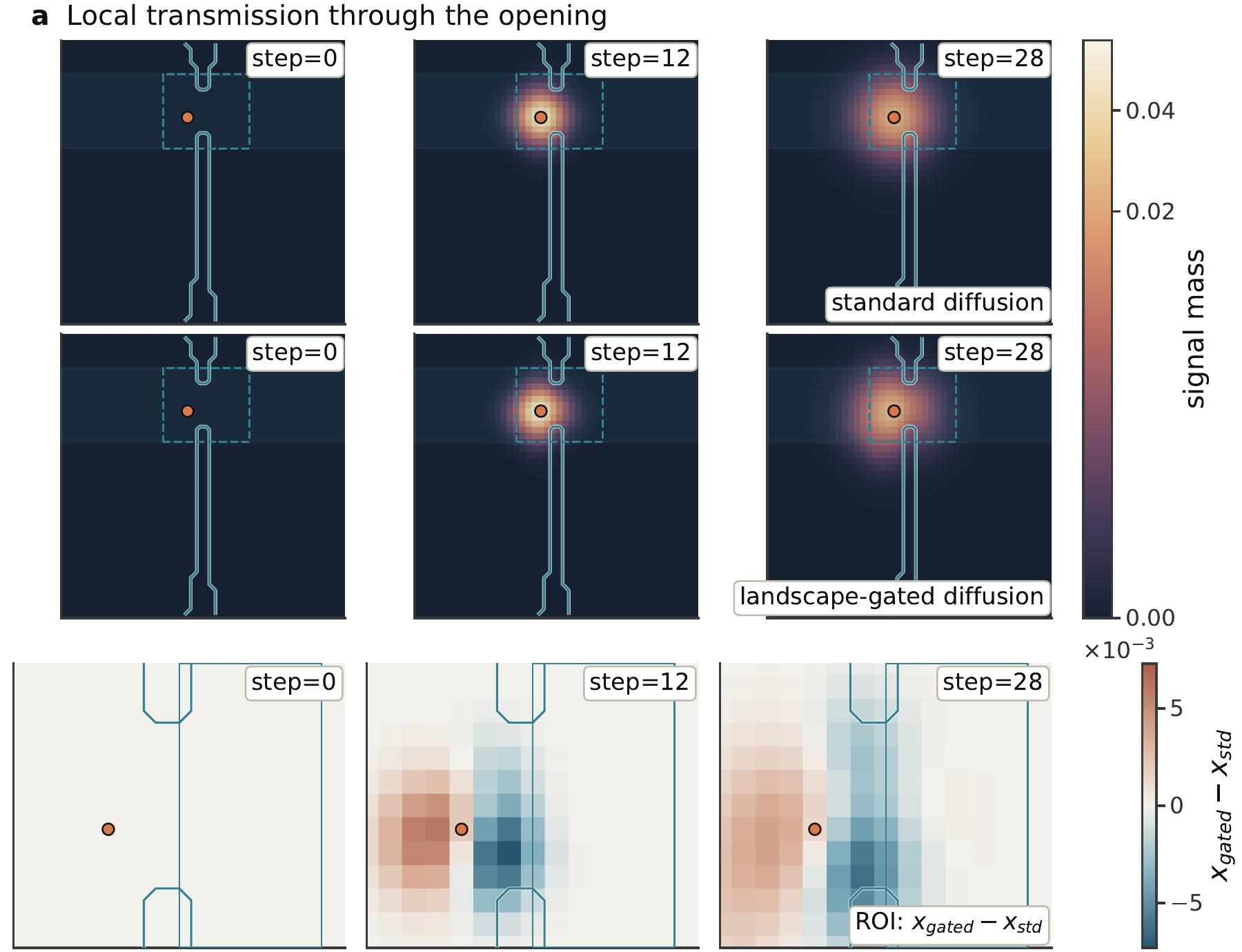}

\vspace{0.8em}

\figimg{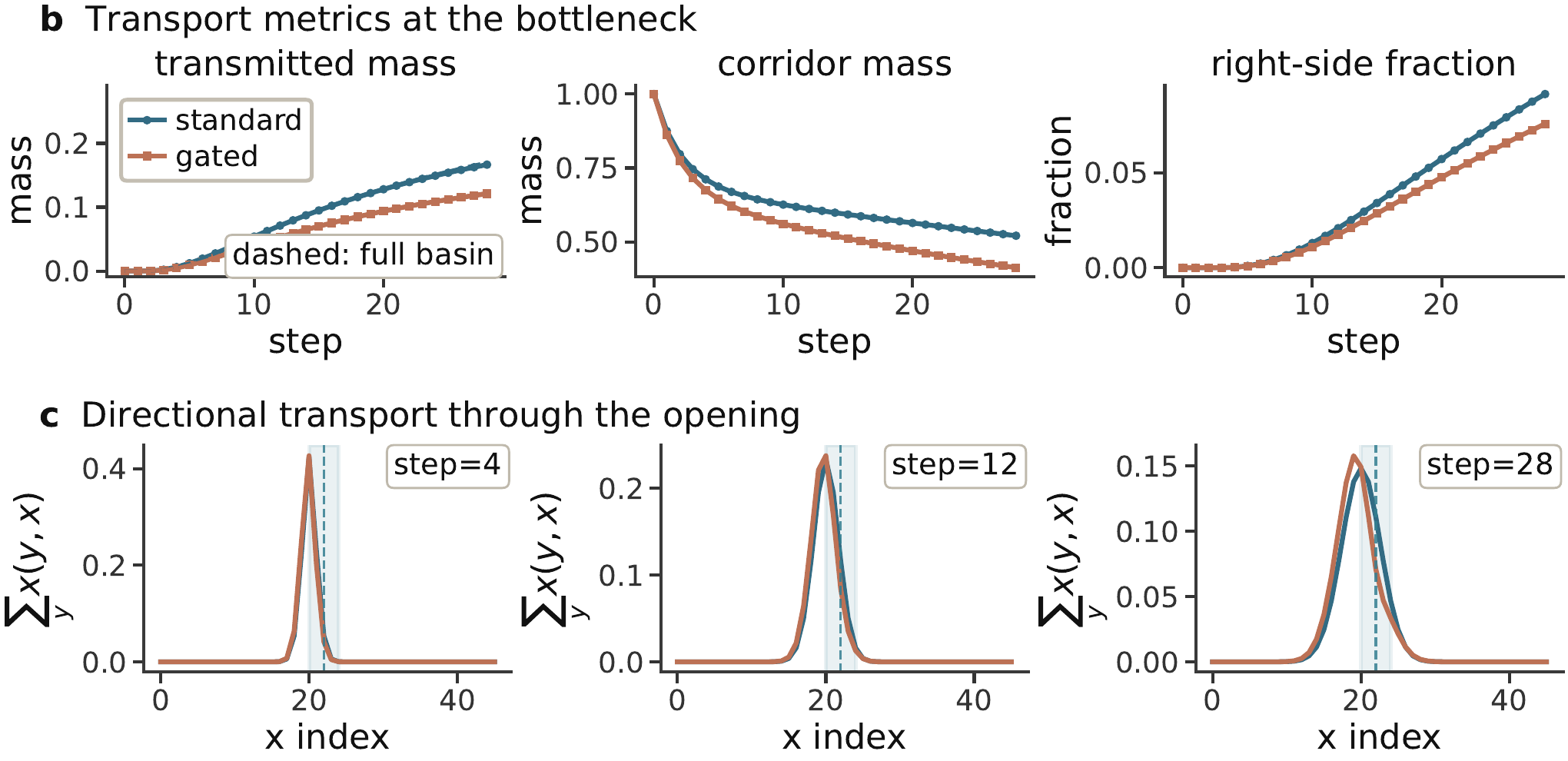}
\caption{Mass-conserving standard and landscape-gated diffusion. (a) Snapshots and local differences at steps 0, 12, and 28. Contours identify the barrier, the shaded band identifies its opening, and the dashed box marks the region of interest. The two signal rows share a nonlinear amplitude scale; the difference row uses a symmetric signed scale. (b) Receiver-window mass (solid), full right-basin mass (dashed), corridor mass, and right-side opening-band fraction. Gating reduces both transmission and corridor mass at step 28. (c) Opening-band horizontal profiles at steps 4, 12, and 28. The same blue and warm series colors denote standard and gated transport throughout.}
\label{fig:diffusion}
\end{figure}

We next consider diffusion across the opening in the central barrier. If \(W\) denotes the nonnegative adjacency weights, the gated weights are \(W^{\mathrm{gated}}_{ij}=W_{ij}\exp(-5c_{ij})\). Each weight matrix is independently row-normalized to a stochastic matrix \(P\). We evolve a unit-mass point source with the lazy push-forward
\begin{equation}
x^{(n+1)}=P_{\mathrm{lazy}}^{\mathsf T}x^{(n)},\qquad
P_{\mathrm{lazy}}=\tfrac12 I+\tfrac12 P.
\label{eq:lazy_diffusion}
\end{equation}
The transpose conserves total mass, and the lazy step removes the alternating parity artifact of a nearest-neighbor bipartite grid. Total mass is one in both runs to numerical precision at every step. The comparisons in Fig.~\ref{fig:diffusion}(a) show that the resulting gating effect is concentrated near the opening-centered bottleneck. The horizontal coordinate in all transport plots is the discrete step, not a calibrated physical time.

At step 28, the receiver-window mass decreases from \(0.166619\) under standard diffusion to \(0.121045\) under gating; the full right-basin mass decreases from \(0.170081\) to \(0.122452\). The opening-corridor mass also decreases, from \(0.520910\) to \(0.412616\), so the corrected experiment does not support increased corridor retention. Instead, retention in the full source basin increases from \(0.829919\) to \(0.877548\). The right-side fraction within the opening band decreases from \(0.091449\) to \(0.075769\). The profiles in Fig.~\ref{fig:diffusion}(c) resolve this redistribution spatially. Gating is therefore a bottleneck-sensitive transport prior whose effect depends on the measurement region.

\subsection{Agmon-biased attention yields block-local interactions}

\begin{figure}[htbp]
\centering
\figimg{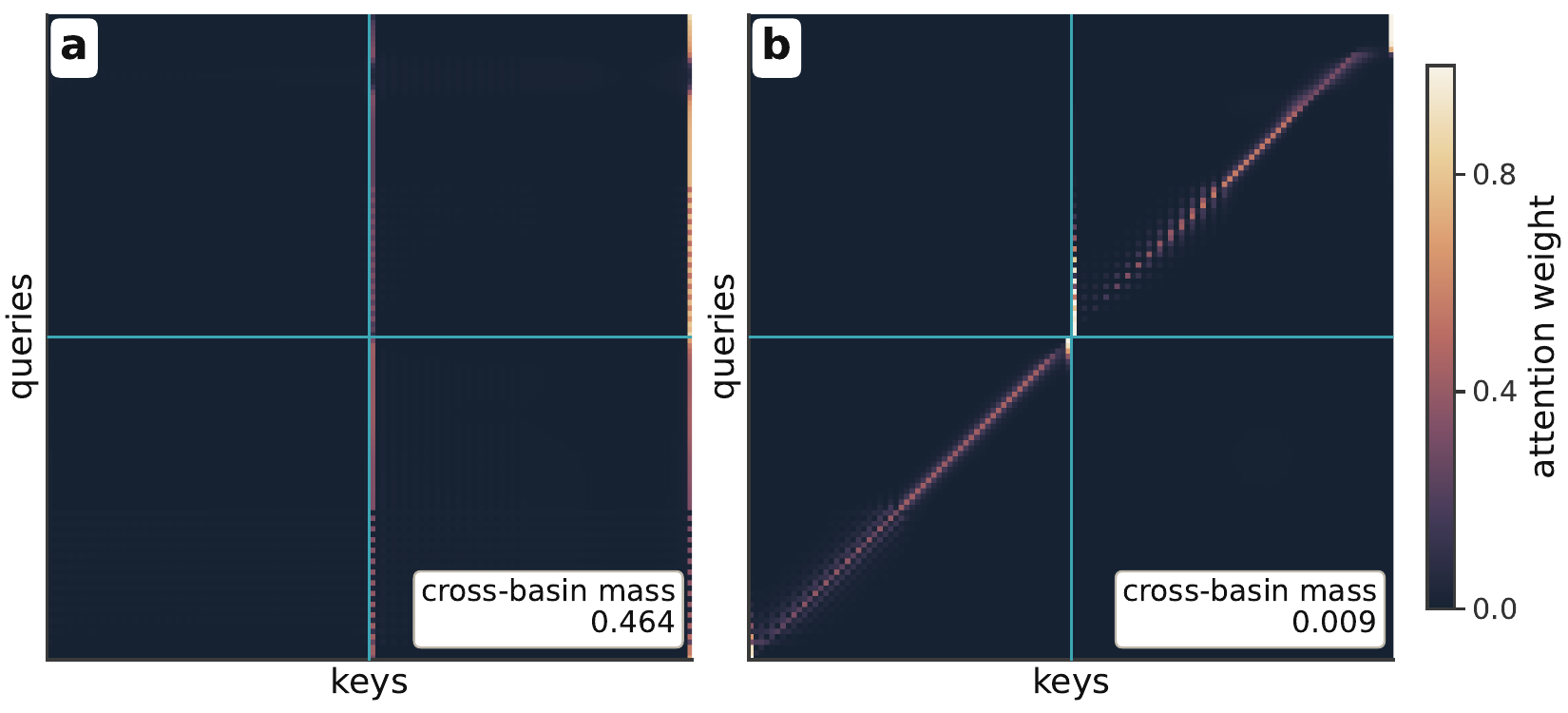}
\caption{Attention locality induced by the discrete Agmon bias. (a) Dense attention has cross-basin mass \(0.463742\). (b) Subtracting \(1.8\rho_E\) from the logits reduces this to \(0.008661\). Both panels use the same node ordering and color scale. Blue-turquoise separators mark basin transitions.}
\label{fig:attention}
\end{figure}

The attention experiment in Fig.~\ref{fig:attention} gives a direct demonstration of the imposed locality bias. We compare feature-based dense attention on 120 nodes with a variant whose logits are penalized by \(1.8\rho_E\). Before applying the fixed query/key projections, each centered eigenmode feature is oriented so that its largest-magnitude entry is nonnegative, removing the arbitrary eigenvector-sign dependence. After sorting nodes by basin, the dense matrix has cross-basin attention mass \(0.463742\), whereas the biased matrix has mass \(0.008661\). The separators show the same basin partition in both panels. This is an illustration of controlled interaction locality, not a test of trained predictive accuracy.

This is precisely the kind of inductive bias one would want when long-range interactions should respect energetic separation rather than raw graph proximity. The result is both interpretable and mathematically well aligned with the underlying theory: nodes separated by large values of the effective barrier should not interact as freely as nodes that are only superficially close in Euclidean or graph distance.

\subsection{Geometry-guided sparsification is mixed, not uniformly beneficial}

\begin{figure}[htbp]
\centering
\figimg{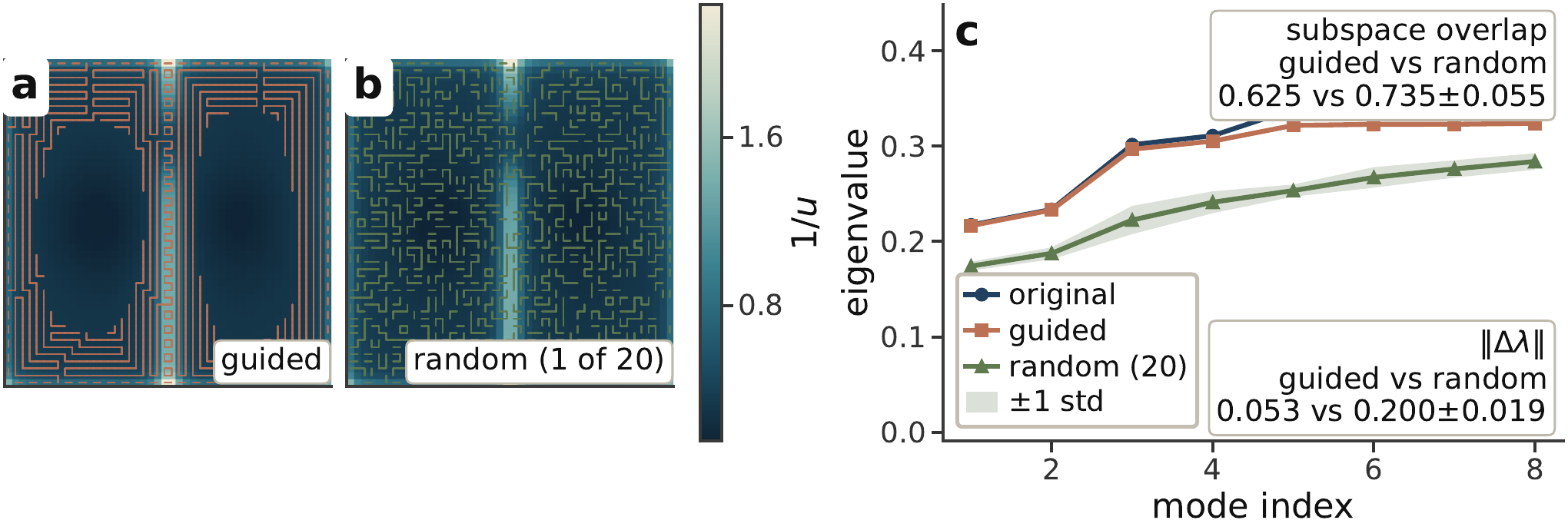}
\caption{Degree-constrained removal of 30\% of edges. (a) Guided pattern. (b) One representative random pattern; statistics use 20 seeds. Both backgrounds show \(1/u\) on the same scale. (c) First eight eigenvalues of the original, guided, and random-pruned operators; shading is one standard deviation of the random spectrum. Guided pruning has smaller eigenvalue error (\(0.053\) versus \(0.200\pm0.019\)) but lower subspace overlap (\(0.625\) versus \(0.735\pm0.055\)).}
\label{fig:sparsify}
\end{figure}

The sparsification experiment in Fig.~\ref{fig:sparsify} is not uniformly positive. Thirty percent of grid edges are removed by a landscape-guided rule or by degree-constrained random pruning, always maintaining a minimum degree of two. The random baseline uses seeds 0--19. Over the first eight eigenvalues, the guided error \(\|\Delta\lambda\|_2\) is \(0.053316\), compared with \(0.200192\pm0.019428\) for random pruning. The shaded spectral band and the reported spread denote one population standard deviation over these 20 draws, not a confidence interval.

The subspace comparison goes in the opposite direction. We measure overlap by the mean singular value of \(Q_0^{\mathsf T}Q_1\), where the columns span the respective first eight eigenspaces; larger values indicate better agreement. Guided pruning gives \(0.624701\), below the random mean \(0.735196\pm0.055127\). Thus a smaller eigenvalue error does not imply better preservation of the associated eigenspace. This distinction is essential to the task-dependent interpretation of the geometry.

\section{Heterogeneous reaction--diffusion example}

To test whether the same framework extends beyond a regular-grid discrete operator, we construct a strongly compartmentalized reaction--diffusion example on a finite-element mesh. The domain contains three Dirichlet cooling holes, several ultra-low-conductivity barrier strips, multiple interior chambers, and a localized hotspot. Barriers combine very low conductivity with large reaction, while chambers combine higher conductivity with weak reaction, producing a medium in which transport is sharply obstructed and confined. The problem is solved in DOLFINx on a Gmsh-generated mesh using the heterogeneous operator
\begin{equation}
-\nabla \cdot \bigl(k(x)\nabla T(x)\bigr) + \sigma(x)\,T(x) = q(x).
\label{eq:heterogeneous_example}
\end{equation}

We use continuous piecewise-linear finite elements with \(17347\) free degrees of freedom. Low modes solve the generalized eigenproblem
\begin{equation}
A_{ff}\phi_j=\lambda_j M_{ff}\phi_j,\qquad
\phi_j^{\mathsf T}M_{ff}\phi_\ell=\delta_{j\ell},
\label{eq:fe_generalized_eigenproblem}
\end{equation}
where \(M_{ff}\) is the assembled mass matrix. The first three eigenvalues in PDE units are \(2.774102\), \(2.989643\), and \(5.045147\). They are comparable to the well threshold \(E=2.238719\) and \(\min(1/u)=2.175715\); in particular, the computed \(\lambda_1\) satisfies the landscape lower-bound diagnostic. Eigenvalues of the stiffness matrix alone would have discretization-dependent matrix scaling and are not used as PDE eigenvalues here.

From the same operator we compute both the physical field \(T\) and the landscape \(u\), then define the effective potential \(1/u\). Wells are extracted automatically from low-\(1/u\) regions after excluding a thin layer near the Dirichlet holes, and basins are assigned by nearest-well Agmon distance. This construction yields sharp wells, a clear basin partition, and low modes that align with the same operator-induced compartments, making the example a clean testbed for our localization framework.

\begin{figure}[htbp]
\centering
\figimg{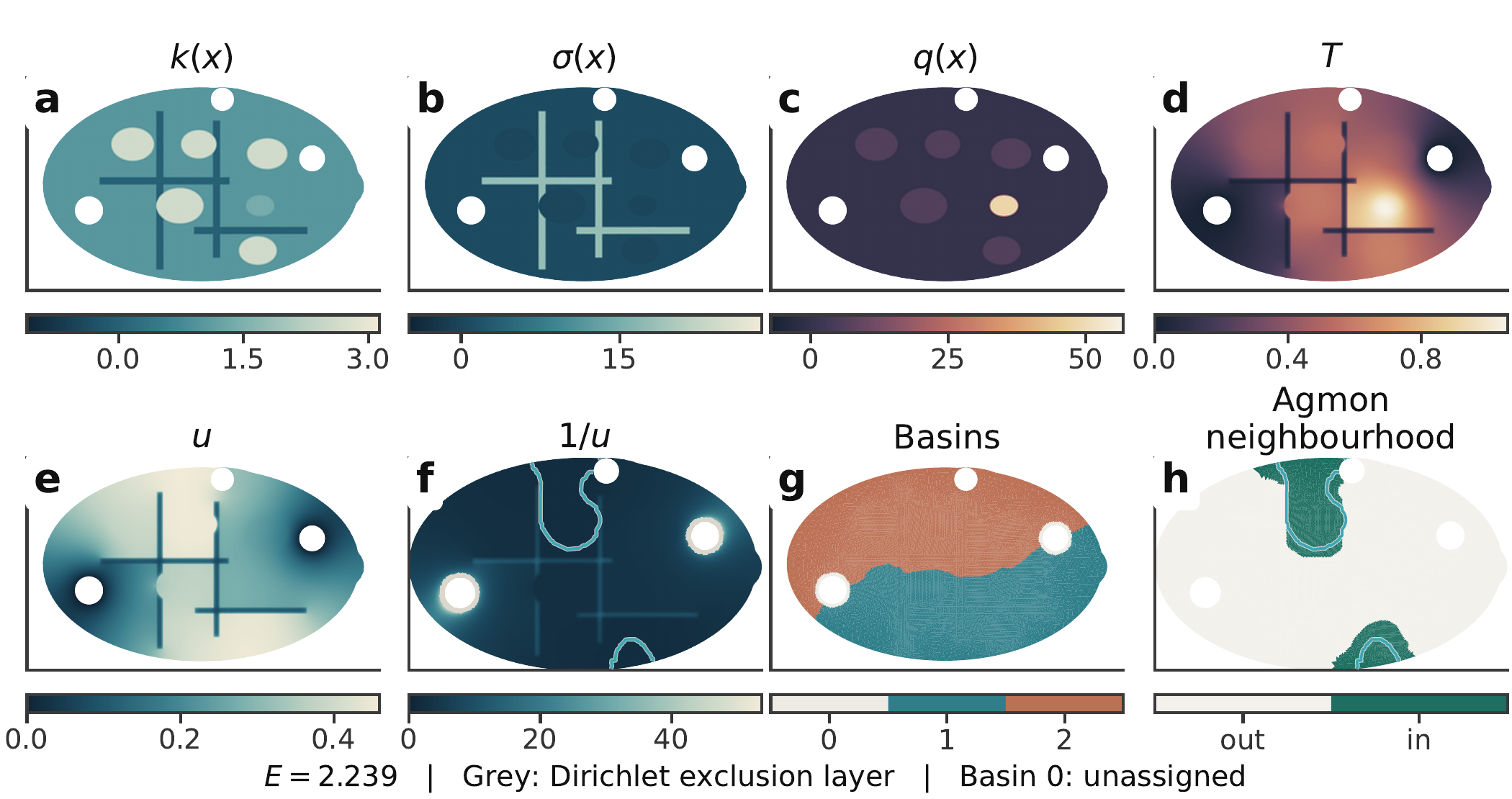}
\caption{Heterogeneous reaction--diffusion geometry. (a--d) Conductivity \(k\), reaction \(\sigma\), source \(q\), and solution \(T\). Coefficients are projected to the continuous visualization space; the PDE uses the original piecewise fields. (e) Landscape \(u\). (f) Effective potential \(1/u\), with well boundaries selected at \(E=2.239\). Grey marks the excluded layer near Dirichlet holes. (g) Discrete basin assignment; 0 denotes unassigned nodes. (h) The neighborhood within graph distance 1.10 of the union of all selected wells. Every panel has its own quantitative or categorical scale. The outer boundary has zero flux; the three holes carry homogeneous Dirichlet data.}
\label{fig:rd}
\end{figure}

The resulting fields confirm that the example produces the intended compartmentalized operator geometry. The heterogeneous coefficients already suggest this structure: the low-conductivity barrier network, spatially varying reaction, and localized source create a solution that is confined to a small subset of the domain rather than spreading uniformly, as seen in Fig.~\ref{fig:rd}(a--d). This same organization reappears in the landscape-based quantities. The landscape \(u\) is smooth at the field level, but the effective potential \(1/u\) sharpens the separation between accessible and shielded regions, producing distinct low-potential wells separated by high barriers. The basin partition and the Agmon neighborhood then make this geometry explicit by identifying the operator-defined compartments and the limited regions that remain effectively connected to each well, as shown in Fig.~\ref{fig:rd}(e--h).

The low modes are consistent with the same picture. Rather than extending broadly across the full domain, the first eigenfunctions are shaped by the barrier network and concentrate within the compartments identified by the effective wells. When their amplitudes are viewed over \(1/u\), the alignment becomes clear: the dominant support of each mode lies in low-\(1/u\) regions, while high-\(1/u\) barriers and the Dirichlet holes suppress cross-compartment communication. This is precisely the behavior the framework is meant to capture. The example therefore shows that the landscape, wells, basins, and Agmon geometry are not merely descriptive visualizations, but practical surrogates for the localization structure that governs the low-energy response of the operator.

The continuum example is also computationally relevant. It shows that the framework carries over naturally to finite-element discretizations built with standard tools such as Gmsh and DOLFINx, and therefore can be integrated into operator-aware numerical workflows without requiring an exotic software stack \cite{Geuzaine2009,Baratta2023}.

\begin{figure}[htbp]
\centering
\figimg{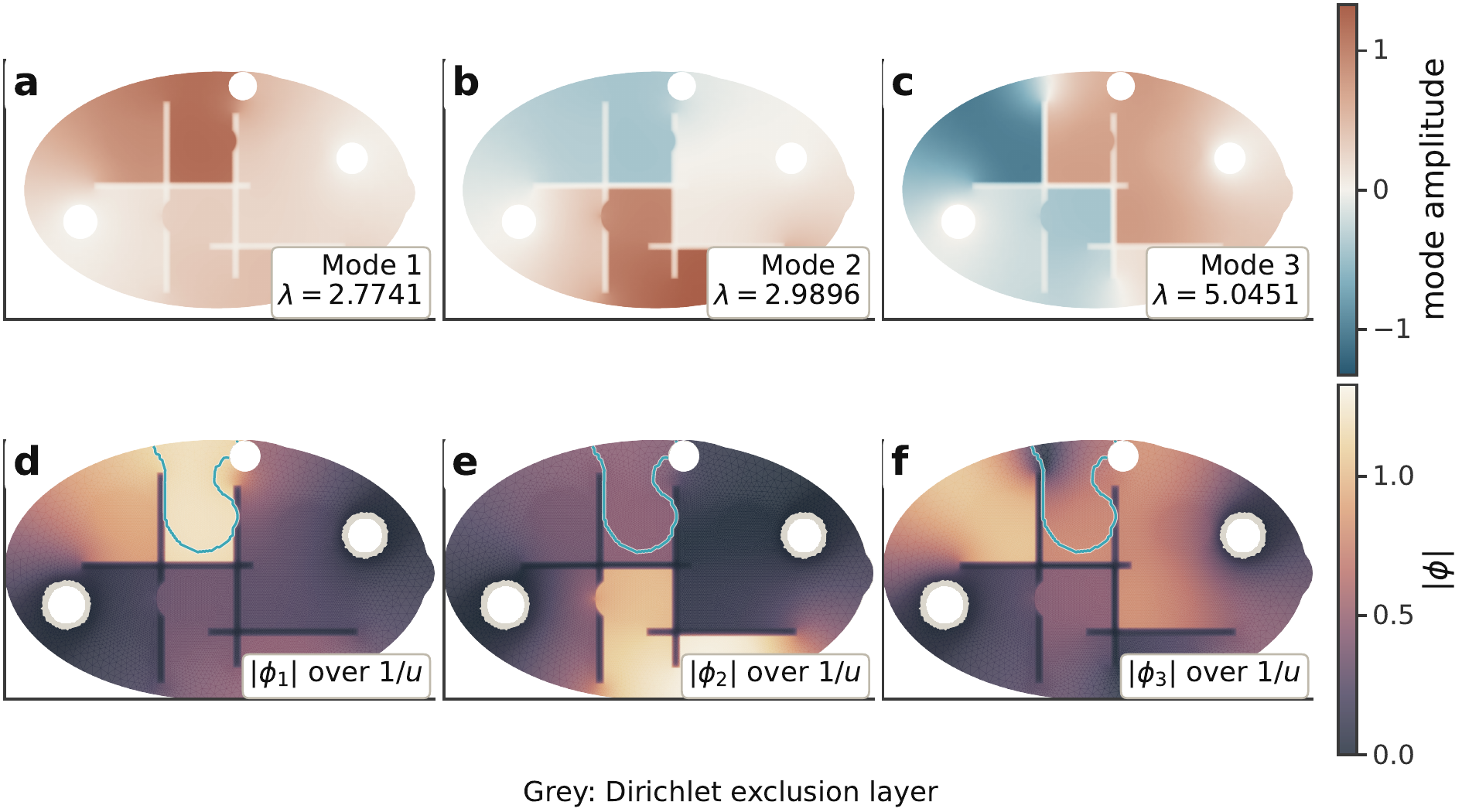}
\caption{Mass-normalized finite-element eigenmodes solving \(A\phi=\lambda M\phi\). (a--c) First three signed modes, with \(\lambda_1=2.7741\), \(\lambda_2=2.9896\), and \(\lambda_3=5.0451\), on a shared symmetric scale. (d--f) Their magnitudes over a neutral \(1/u\) underlay, with a shared amplitude scale and blue-turquoise well contours. Grey indicates the Dirichlet exclusion layer. These PDE eigenvalues, rather than stiffness-only matrix eigenvalues, are the relevant spectral quantities.}
\label{fig:rdmodes}
\end{figure}

\section{Conclusions}

This study supports a clear but deliberately qualified conclusion. The localization landscape provides much more than an appealing visualization of heterogeneous operators. Through the effective potential \(\Weff=1/u\), it induces a geometry that identifies wells, barriers, bottlenecks, basin structure, and regions of effective accessibility. In both the discrete and continuum examples studied here, that geometry correctly predicts where low-energy eigenmodes live and how confinement organizes the response of the operator.

The strongest downstream successes occur when the task is intrinsically aligned with low-energy accessibility. This is most evident in three places: the recovery of the low-mode partition, the construction of adaptive Agmon neighborhoods on a fixed graph, and the suppression of spurious cross-basin interactions in attention-like operators. In each of these cases, localization-landscape geometry provides an interpretable and operator-aware notion of locality that is richer than Euclidean proximity or hop distance alone.

At the same time, the framework should not be presented as a universal improvement mechanism. In the diffusion and sparsification experiments, the effects are meaningful but more nuanced: the geometry acts as a bottleneck-sensitive transport prior and a low-energy spectral preservation prior, respectively, rather than as a blanket performance enhancer. The broader methodological lesson is therefore precise. Operator-induced geometry is a powerful inductive bias when relevance is governed by barriers, confinement, and low-energy communication; outside that regime, its transfer can be mixed.

These observations suggest a practical direction for future work. Instead of using localization landscapes only as a diagnostic tool, one can use them as a computational layer that supplies operator-aware neighborhoods, locality priors, sparsification rules, and geometry-aware couplings in PDE solvers, graph algorithms, and learning systems. In that role, the landscape is not merely descriptive. It becomes a tractable surrogate for the hidden geometry that actually governs localization, propagation, and interaction in heterogeneous operators.

\FloatBarrier

\paragraph{AI-assisted writing}
OpenAI assistants were used for language editing, notebook refactoring, debugging figure generation, and consistency checks between computed results and the manuscript. The numerical results were produced by the documented scientific code. The author is responsible for reviewing the final content and all scientific claims before submission.

\paragraph{Code availability}

The revised notebook and manuscript sources are maintained in the private GitHub repository \href{https://github.com/jgavila01/project-landscape}{jgavila01/project-landscape}. Access may be requested from the author. The notebook regenerates the figures and numerical diagnostics; this revision does not claim that the corrected code is publicly archived.

\FloatBarrier

\end{document}